\documentclass[conference]{IEEEtran}  % final version
\usepackage{ifpdf}
\ifCLASSINFOpdf
   \usepackage[pdftex]{graphicx}
\else
    \usepackage[dvips]{graphicx}
\fi
\usepackage[cmex10]{amsmath}
\usepackage{amssymb}
\usepackage{amsthm}
\usepackage{cite}
\usepackage{url}
\usepackage{bm}
\usepackage{subfigure}
\usepackage[usenames,dvipsnames]{color}

\newcommand{\bmrm}[1]{\bm{{\rm #1}}}
\newcommand{\Pm}[2]{\bmrm{P}'^{(#1)}_{#2}}
\newcommand{\etal}{{\it et al.}~}
\newcommand{\T}{{\mathsf{T}}}
\newcommand{\mlh}{m^{(l)}_{h}}
\newcommand{\vlh}{v^{(l)}_{h}}
\newcommand{\bmlh}{\bar{m}^{(l)}_{h}}
\newcommand{\bvlh}{\bar{v}^{(l)}_{h}}

\IEEEoverridecommandlockouts   %allows using \thanks

\begin{document}
\title{
Threshold Improvement of \\ Low-Density Lattice Codes via Spatial Coupling
}
\author{
\authorblockN{Hironori Uchikawa\authorrefmark{1}, Brian M.~Kurkoski\authorrefmark{2}, Kenta Kasai\authorrefmark{1} and Kohichi Sakaniwa\authorrefmark{1}\\\\ }
\authorblockA{\authorrefmark{1} Dept. of Communications and Integrated Systems, Tokyo Institute of Technology, 152-8550 Tokyo, Japan.\\Email: \{uchikawa, kenta, sakaniwa\}@comm.ss.titech.ac.jp} 
\authorblockA{\authorrefmark{2} Dept. of Information and Communications Engineering, University of Electro-Communications, 182-8585 Tokyo, Japan. \\Email: kurkoski@ice.uec.ac.jp} \\
 \thanks{B.K.~was supported in part by the Ministry of Education, Science, Sports and Culture; Grant-in-Aid for Scientific Research (C) number 21560388 and Grant-in-Aid for Scientific Research (C) number 23560439. } }
% \author{
%  \IEEEauthorblockN{Hironori Uchikawa, Kenta Kasai and Kohichi Sakaniwa}
%  \IEEEauthorblockA{Dept.\ of Communications and Integrated
%  Systems\\Tokyo Institute of Technology\\152-8550 Tokyo, JAPAN\\
%  Email: \{uchikawa, kenta, sakaniwa\}@comm.ss.titech.ac.jp}
%  \and
% \IEEEauthorblockN{Brian M.~Kurkoski} 
%  \IEEEauthorblockA{University of Electro-Communications\\ 
%  Tokyo, Japan\\ 
%  Email: kurkoski@ice.uec.ac.jp}
% }



\maketitle

\begin{abstract}
%\boldmath
% 既存研究と自分の研究の違いを明確に書く
Spatially-coupled low-density lattice codes 
(LDLC) are constructed using protographs. 
Using Monte Carlo density evolution using single-Gaussian messages, 
we observe that the threshold of the spatially-coupled LDLC 
is within 0.22 dB of capacity of the unconstrained power channel.   
This is in contrast with a 0.5 dB noise threshold for the conventional LDLC lattice construction.

\end{abstract}


% \fi
\IEEEpeerreviewmaketitle

\newcommand{\bmk}[1]{{\color{Mahogany} #1}}  
\newcommand{\umk}[1]{{\color{PineGreen} #1}} % edited by uchikawa

\section{Introduction}

Kudekar \etal rigorously proved that
the belief-propagation (BP) threshold, 
the maximum channel parameter (worst channel condition) 
that allows transmission with an arbitrary small error probability,
for low-density parity-check
(LDPC) codes improves up to the maximum-a-posteriori (MAP) threshold
by spatial coupling.
This phenomenon is called {\it threshold saturation} \cite{kudekar2011it}.
The threshold saturation phenomenon is observed 
not only for the binary erasure channel (BEC), 
but also for general binary memoryless symmetric
channels \cite{kudekar2010bms}. Moreover empirical evidence
via density evolution analysis has been done for other channels, 
such as the multiple access channel \cite{kudekar2011isit-mac}, 
a relay channel \cite{uchi2011isit}, and a channel with memory
\cite{kudekar2011isit-mem}.

The performance improvement via spatial coupling has not only been
reported for LDPC codes, but also many other problems.
For example, compressed sensing \cite{kudekar2010cs},
BP-based multiuser detection for 
randomly-spread code-division multiple-access (CDMA)
\cite{takeuchi2011isit}, and K-SAT problem \cite{hassani2010itw}
have all shown a benefit by using spatial coupling.
We conjecture that the threshold saturation phenomenon is 
universal for graphical models, in particular for sparse
systems.

Low-density lattice codes (LDLC) are lattices defined 
by a sparse inverse generator matrix.
Sommer, Feder and Shalvi proposed this lattice construction,
described its BP decoder, and gave extensive
convergence analysis \cite{sommer2008it}. 
Since decoding complexity is linear in the dimension, 
it is possible to decode LDLC lattices with dimension of $10^6$. 
Although LDLC lattices can be decoded efficiently, capacity-achieving 
LDLC lattices have not so far been constructed.
The best-known result is that a noise threshold of LDLC 
appeared within 0.6 dB of the capacity of the 
unconstrained-power AWGN channel \cite{sommer2008it}.

In this paper, we consider spatially-coupled LDLC, which will be abbreviated SC-LDLC.
By using Monte Carlo density evolution with a single-Gaussian 
approximation, it is observed that the threshold of SC-LDLC approaches the theoretical limit within 0.22 dB.

\section{Low Density Lattice Codes and Their Protographs}

\subsection{Lattices}
An $n$-dimensional lattice $\Lambda$ is defined by an $n$-by-$n$
generator matrix $\bmrm{G}$. The lattice consists of the discrete set of
points $\bmrm{x} = (x_1,x_2, \ldots, x_n)^\T$ for which
\begin{align*}
\bmrm{x} = \bmrm{G} \bmrm{b}, 
%\label{eqn:lattice}
\end{align*}
where $\bmrm{b} =(b_1,\ldots, b_n)^\T$ is from the set of all possible
integer vectors, $b_i \in \mathbb Z$.  The transpose of a vector $\bmrm x$ is denoted $\bmrm x ^\T$.
Lattices are linear, in the sense that 
$\bmrm{x}_1 + \bmrm{x}_2 \in \Lambda$
if $\bmrm{x}_1$ and $\bmrm{x}_2$ are lattice points.
It is assumed that $\bmrm{G}$ is
$n$-by-$n$ and full rank (note some definitions of SC-LDLC allow 
$\bmrm{G}$ to have  additional rows which are linearly
dependent).  

We consider the unconstrained power system, as was used
by Sommer \etal \cite{sommer2008it}. 
Let the codeword $\bmrm{x}$ be an arbitrary
point of the lattice $\Lambda$. This codeword is transmitted over
an AWGN channel, where noise $z_{i}$ with noise variance $\sigma^2$
is added to each code symbol. Then the received sequence
$\bmrm{y} = (y_1, y_2, \ldots, y_n)$ is
$y_i = x_{i} + z_i,$ for $i = 1,2,\ldots, n$.  A maximum-likelihood 
decoder selects $\hat{\bmrm{x}}$ as the estimated
codeword, and a decoder error is declared if $\hat{\bmrm{x}} \neq \bmrm{x}$.
The capacity of this channel is the maximum noise power
at which a maximum-likelihood decoder can recover the
transmitted lattice point with error probability as low as
desired. In the limit that $n$ becomes asymptotically large, there
exist lattices which satisfy this condition if and only if
\cite{Po1994it}:
\begin{align}
 \sigma^{2} \leq \frac{|\rm{det}(\bmrm{G})|^{2/{\it n}}}{2 \pi e}.
 \label{eqn:cap}
\end{align}
In the above $|\rm{det}(\bmrm{G})|$ is the volume of the Voronoi region,
which is the measure of lattice density.

\subsection{Low Density Lattice Codes}
An LDLC is a dimension $n$ lattice with 
a non-singular generator matrix $\bmrm{G}$. An inverse generator matrix
$\bmrm{H} = \bmrm{G}^{-1}$ of the LDLC is sparse, so that LDLC can be decoded 
using BP \cite{sommer2008it}. 
The $n$-by-$n$ matrix $\bmrm{H}$ of an ($\alpha$, $d$) LDLC has row and column weight $d$, where each row and column has one non-zero entry of weight $\pm 1$ and $d-1$ entries with weight which depends upon $\alpha$. More precisely, the matrix $\bmrm{H}$ is defined
as: 
\begin{align}
 \bmrm{H} = \bmrm{P}'_{1} + w \sum_{i=2}^{d} \bmrm{P}'_{i},
\label{eqn:matH}
\end{align}
where 
\begin{align}
\bmrm{P}'_{i} = \bmrm{S}_{i} \bmrm{P}_{i}. 
\label{eqn:Pprime}
\end{align}
$\bmrm{S}_{i}$ denotes a random sign change matrix, 
$\bmrm{P}_{i}$ denotes a random permutation matrix, and 
\begin{align*}
 w = \sqrt{\frac{\alpha}{d-1}}.
\end{align*}
We choose $0 \leq \alpha \leq 1$, so that BP decoding of LDLC will 
converge exponentially fast \cite{sommer2008it}.
The permutations result in $\bmrm{H}$ having exactly one 1 and 
exactly $d-1$ $w$'s in each column and row. The random sign change matrix
$\bmrm{S}_{i}$ is a square, diagonal matrix, where the diagonal entries are
$+1$ or $-1$ with probability 1/2. The bipartite graph of an LDLC can be 
defined similarly to LDPC codes \cite{sommer2008it}.

For the lattice construction considered in this paper, each row and each column of $\bmrm H$ has one 1 and $d-1$
$w$'s, and with $w  \ll  1$ the power is suitably normalized, since such lattices
have $|\rm{det}(\bmrm{G})| \rightarrow 1$ as the dimension becomes large.

\subsection{Protograph of LDLC}

\begin{figure}[t]
  \begin{center}
   \includegraphics[scale=0.55]{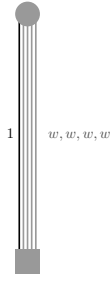}
   \caption{Protograph of ($\alpha$, $d=5$) LDLC.
   The circle and rectangle nodes represent 
   variable and check nodes, respectively.
   Black line denotes a edge labeled 1 and gray lines denote edges labeled $w$.}
   \label{fig:ldlc}
  \end{center}
 \end{figure}

As with protograph-based LDPC codes, it is possible to construct a
matrix $\bmrm{H}$ from a protograph. 
Fig.~\ref{fig:ldlc} shows a protograph of an ($\alpha$, $d$) LDLC for $d=5$.
%(see \cite{thorpe-JPL2003} for the definition of protographs).
The circle and rectangle nodes represent 
variable and check nodes, respectively. 
The black edge denotes the label 1 edge and gray edges denote the label $w$ edges. 
The degree of each node  is $d=5$.
From the protograph, the usual bipartite graph of the LDLC is generated by 
a {\it copy-and-permute} operation \cite{thorpe-JPL2003}, 
with random sign changes and label assignments for respective edges.

% $M$ を使ってM次元のLDLCを構成する際の話をここで述べておく
%Let $M$ denote the number of both variable and check nodes.
%For our example, $M=100$ means that we have 100 copies of
%the protograph so that we have 100 variable nodes, 100 check nodes, 
%100 label 1 edges, and $100(d-1)$ label $w$ edges. 
%Edges are permuted, so that.

\section{Spatially Coupled Protograph of LDLC}
In this section, we first define ($\alpha, d, L$)
SC-LDLC, then ($\alpha, d, L, K$) SC-LDLC will be introduced.

\subsection{Standard Coupling}
We define a ($\alpha,d,L$) SC-LDLC as a dimension $n(L-d+1)$
lattice with an $n(L-d+1) \times nL$ inverse generator matrix 
$\bmrm{H}_{[L]} $ 
as described by Eq.~\eqref{eqn:HadL}.
The structure of $\bmrm{H}_{[L]}$ is similar to 
the parity check matrix of tail-biting convolutional codes.
In Eq.~\eqref{eqn:HadL}, 
$\bmrm{H}^{(l)} = \Pm{l}{1} + w \sum_{i=2}^{d} \Pm{l}{i}$ is 
an inverse generator matrix of a $n$ dimension ($\alpha,d$) LDLC for $1 \leq l \leq L$, and each $\Pm{l}{d}$ represents a distinct matrix of the form of (\ref{eqn:Pprime}), for distinct $l$ and $d$.

In this construction, some integers are terminated to 0.   The integer vector of the form:
\begin{align*}
\bmrm{\tilde{b}} &= 
\begin{bmatrix}
\bmrm{b} \\
\bmrm{0}_{n(d-1)}
\end{bmatrix},
\end{align*}
is used, so that if $\bmrm{x} \in \Lambda$, then $\bmrm{H}_{[L]} \cdot \bmrm{x} = \bmrm{\widetilde b}$.  Here,  $\bmrm{b} =(b_1,\ldots, b_i, \ldots, b_{n(L-d+1)})^\T$ is an information integer
vector, i.e. $b_i \in \mathbb Z$.  And, $\bmrm{0}_{n(d-1)}$ is the all zero column vector of length $n(d-1)$.  The dimension of the lattice is, therefore, less than the number of elements in $\bmrm x$, which is $nL$.

\begin{figure}[t]
  \begin{center}
   \includegraphics[scale=0.5]{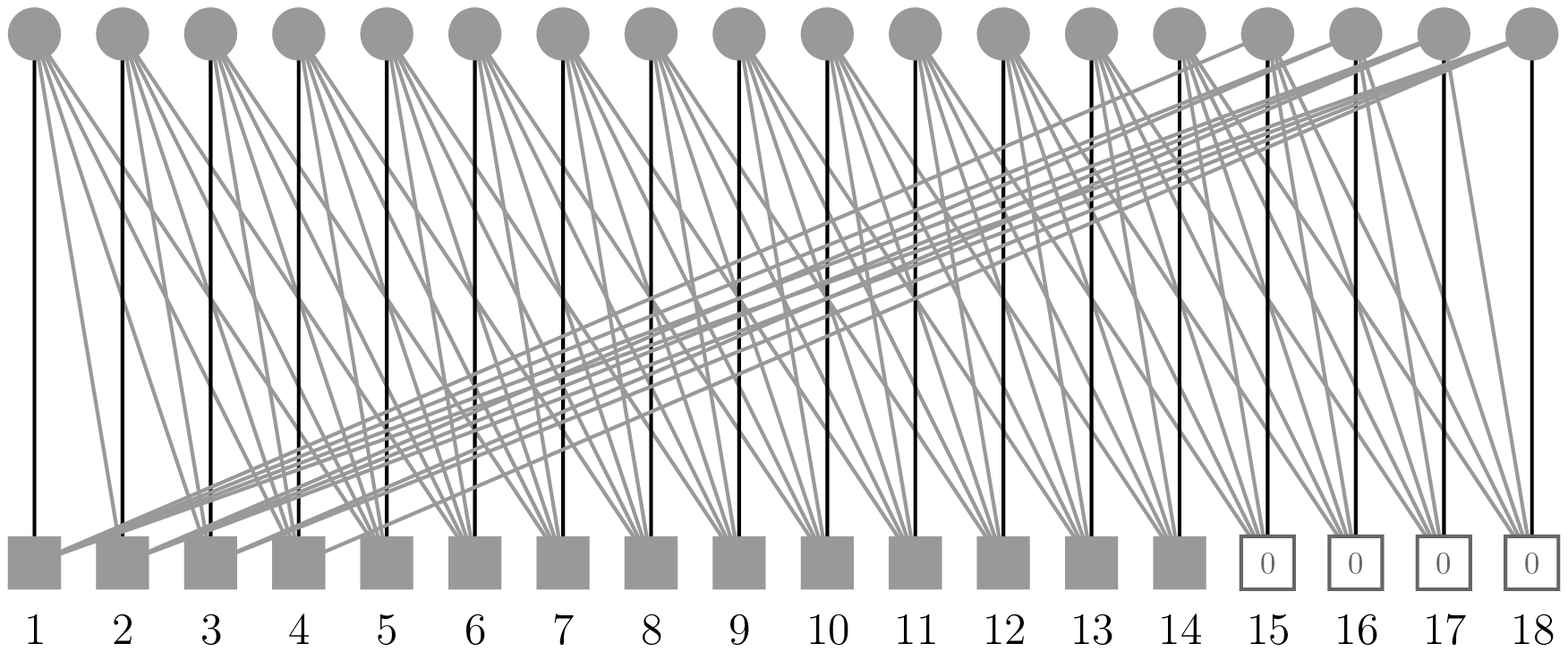}
   \caption{Protograph of ($\alpha$, $d=5$, $L=18$) SC-LDLC. White rectangle
   nodes are null check nodes corresponding to $\bmrm{0}_{n(d-1)}$. }
   \label{fig:ldlc-shorten}
  \end{center}
 \end{figure}

The protograph of ($\alpha$, 5, 18) SC-LDLC is shown in Fig.~\ref{fig:ldlc-shorten}.
Reliable messages from white rectangle nodes (null check nodes), 
i.e., 15, 16, 17, and 18, are gradually propagated.
In the protograph, a node corresponds to $n$ variable nodes or $n$ check nodes, and nodes are connected according to an entry in  Eq.~\eqref{eqn:HadL} corresponding to that edge.
BP decoding, as well as density evolution, proceeds on the protograph,
using $\bmrm{H}_{[L]}$, as with protograph-based LDPC codes \cite{thorpe-JPL2003}.

Associated with the lattice is an $nL \times n(L-d+1)$ generator matrix $\bmrm{G}_{[L]}$. The $\bmrm{G}_{[L]}$ has a sub-matrix from column 1 to
$n(L-d+1)$ which is an $nL \times nL$ non-singular matrix $\bmrm{\tilde{G}}_{[L]}$.
Therefore a lattice point in the dimension $n(L-d+1)$ lattice is generated with
\begin{align}
\bmrm{x} = \bmrm{G}_{[L]} \bmrm{b} = \bmrm{\tilde{G}}_{[L]} \bmrm{\tilde{b}}.
\label{eqn:sclattice}
\end{align}

Dimension ratio is defined as
\begin{align*}
R_{L} = \frac{n(L-d+1)}{nL} = 1 - \frac{d-1}{L}.
\end{align*}
The ratio $R_{L}$ converges to 1 with increasing $L$, with gap $O(1/L)$. 
Therefore, this dimension loss is negligible for sufficiently large $L$.

%%%%%%%%%%%%%%%%%%%%%%%%%%%%%%%%%%%%%%%%%%%%%%%%%%%%%%%%%%%%%%%%%%%%%%%%%%%%%%%%%%%%%%%%%%%%%%%%%%%%%%%%%%%%%%
% tail biting matrix
%%%%%%%%%%%%%%%%%%%%%%%%%%%%%%%%%%%%%%%%%%%%%%%%%%%%%%%%%%%%%%%%%%%%%%%%%%%%%%%%%%%%%%%%%%%%%%%%%%%%%%%%%%%%%%
\begin{figure*}[ht]
\begin{align}
 \bmrm{H}_{[L]} &= 
  \begin{bmatrix}
   \Pm{1}{1}  &              &        &        &               & w\Pm{L-d+2}{d} & \cdots & w\Pm{L}{2} \\
   w\Pm{1}{2} & \Pm{2}{1}    &        &        &               &                & \ddots & \vdots  \\
   \vdots     & \vdots       & \ddots &        &               &         &  & w\Pm{L}{d}  \\
   w\Pm{1}{d} & w\Pm{2}{d-1} & \ddots & \ddots \\
              & \ddots       & \ddots & \ddots & \Pm{L-d+1}{1} \\
              &              & \ddots & \ddots & w\Pm{L-d+1}{2} & \Pm{L-d+2}{1} \\
              &              &        & \ddots & \vdots & \vdots & \ddots \\
              &              &        &        & w\Pm{L-d+1}{d} & w\Pm{L-d+2}{d-1} & \cdots & \Pm{L}{1}
  \end{bmatrix}
\label{eqn:HadL}
\end{align}
\hrulefill
\end{figure*}
%%%%%%%%%%%%%%%%%%%%%%%%%%%%%%%%%%%%%%%%%%%%%%%%%%%%%%%%%%%%%%%%%%%%%%%%%%%%%%%%%%%%%%%%%%%%%%%%%%%%%%%%%%%%%%

\subsection{Randomized coupling}
In order to simplify the evaluation of the noise threshold, we define a
($\alpha,d,L,K$) SC-LDLC in this section.
A bipartite graph of a code in the $(\alpha, d, L, K)$ SC-LDLC
is constructed as follows.
Similar to the ($\alpha,d,L$) SC-LDLC, for each position $l \in \{1,\ldots,L\}$,
there are $n$ variable nodes and $n$ check nodes.
Each of the $d$ edges of a variable node at 
position $l$ connects uniformly and independently to 
a check node at position $\hat{l} \in \{l,\ldots,l+K-1\}$, 
so that every check node has one edge labeled 1 and $d-1$ edges labeled $w$.
Note that $\hat{l}$ is taken modulo $L$ value, if $\hat{l} > L$.
In the same way, each of the $d$ edges of a check node at 
position $l$ connects uniformly and independently to 
a variable node at position 
$\check{l} \in \{l-K+1,\ldots,l\}$, 
so that every variable node has one label 1 edge and $d-1$ label $w$ edges.
Note that $\check{l}$ is taken modulo $L$.
The variable $K$ is called {\it smoothing} parameter, and $K < L$.
This approach of randomizing connections is based on that of spatially-coupled LDPC codes \cite{kudekar2011it}.
Since $n(K-1)$ check nodes do not have information, the dimension ratio $R_{L}$ of 
the ($\alpha,d,L,K$) SC-LDLC is as follows:
\begin{align*}
R_{L} = \frac{n(L-K+1)}{nL} = 1 - \frac{K-1}{L}.
\end{align*}
Similar to the ($\alpha,d,L$) SC-LDLC, $R_{L}$ converges to 1 as
increasing $L$ with gap $O(1/L)$. 
Therefore we can neglect this dimension loss for sufficient large $L$.
We employ ($\alpha,d,L,K$) SC-LDLC for density evolution
analysis in the next section.

\section{Monte Carlo Density Evolution for \\Single Gaussian Decoder of SC-LDLC}

In this section we describe a method to find noise thresholds for ($\alpha,d,L,K$) SC-LDLC lattices over the unconstrained-power AWGN channel.
For LDPC codes, the BP threshold may be easily evaluated using 
density evolution which tracks the probability density function of 
log likelihood ratio messages exchanged 
between the variable nodes and the check nodes 
in a bipartite graph \cite{richardson01capacity}.
However, density evolution of LDLC lattices is much more complicated.
Since messages exchanged in the BP decoder for LDLC lattices
are probability density functions, density evolution of the BP decoder must track 
the probability density function of the message probability density functions.
Due to the space limitations,  the LDLC lattice BP decoding procedure is omitted; please refer to the paper of Sommer \etal \cite{sommer2008it} for details.

%%%%%%%%%%%%%%%%
Here, noise thresholds are found via density evolution, using not true
BP, but instead a single-Guassian approximation of the decoder message
\cite{brian09sgde}.  In this approximation, the probability density
function is approximated using a single-Gaussian message, which is
described by just two scalars for each message: a mean and a variance.
For conventional LDLC lattices, this approximate method is effective,
giving a noise threshold of 0.6 dB \cite{brian09sgde}, the same as the
0.6 dB waterfall region of a dimension $10^6$ lattice
\cite{sommer2008it}.  Performing density evolution would require a
joint distribution in two scalars. While not intractable, this is
nonetheless computationally demanding. Instead, Monte Carlo density
evolution will be used, as has been done for non-binary low-density
parity check codes \cite{golgo2010mcde}.  We describe Monte Carlo
density evolution for the single-Gaussian decoder of the
($\alpha,d,L,K$) SC-LDLC as follows.

Variable (check) node at each position $l$ has
two message pools, ${\cal P}^{(l)}_{1}$ and ${\cal P}^{(l)}_{w}$ (${\cal \bar{P}}^{(l)}_{1}$ and 
${\cal \bar{P}}^{(l)}_{w}$), 
which distinguish the edges labeled 1 from those labeled $w$. 
Both ${\cal P}^{(l)}_{h}$ and ${\cal \bar{P}}^{(l)}_{h}$
have $N_{\textrm s}$ messages, i.e., 
\begin{eqnarray*}
{\cal P}^{(l)}_{h} &\!\!\!\!=\!\!\!\!\!& \{(\mlh[1], \vlh[1]), \ldots, (\mlh[N_{s}],\vlh[N_{s}])\} \\ 
\Big({\cal \bar{P}}^{(l)}_{h} &\!\!\!\!\!=\!\!\!\!\!& \{(\bmlh[1], \bvlh[1]), \ldots, (\bmlh[N_{s}],\bvlh[N_{s}])\}\Big),
\end{eqnarray*} 
for $h \in \{1, w\}$. 
In the following, the index $[i]$ is omitted. % of a $(\mlh[i], \vlh[i])$ ($(\bmlh[i], \bvlh[i])$) for the simplicity.
The pair ($\mlh, \vlh$) $\left( (\bar{m}^{(l)}_{h}, \bar{v}^{(l)}_{h}) \right)$ 
denotes a mean and a variance of a variable-to-check (check-to-variable)
message transmitted from position $l$ along an edge labeled $h$.

\subsubsection{Initialization}
The messages ($\mlh, \vlh$) $\in {\cal P}^{(l)}_{h}$ for all 
$l \in \{1,\ldots, L\}$ and $h \in \{1,w\}$ is initialized as follows:
The noise variance $\sigma^2$ is assigned to $\vlh$ and 
the received symbol $\mu$ 
generated from ${\cal N}(0, \sigma^2)$ is assigned to $\mlh$,
since the all zero codeword, i.e., $b_{i} = 0$ for all $i \in \{1,\ldots, n(L-K+1)\}$, is assumed.

At each half iteration, $N_{s}$ messages in ${\cal P}^{(l)}_{h}$ and ${\cal \bar{P}}^{(l)}_{h}$
are computed alternately for each label $h$ at each position $l$ in the
following way.
\subsubsection{Check node operation}
The ($\bmlh, \bvlh$) $\in {\cal \bar{P}}^{(l)}_{h}$ for all 
$l \in \{1,\ldots, L\}$ and $h \in \{1,w\}$ is computed by
\begin{align*}
 \bar{m}^{(l)}_{h} = \frac{1}{h} \sum^{d-1}_{j=1}h_{j}
 m^{(l_{j})}_{h_{j}}, \ \ \ \ \ 
 \bar{v}^{(l)}_{h} = \frac{\sum^{d-1}_{j=1}h_{j}^{2} v^{(l_{j})}_{h_{j}} }{h^2}, 
\end{align*}
where $l_{j} \in {l-w+1, \ldots, l} \mod{L}$, and 
$h_{j} \in \{1, w\}$. 
The $d-1$ messages ($m^{(l_{j})}_{h_{j}}, v^{(l_{j})}_{h_{j}}$) are chosen as follows:
first the position $l_{j}$ is uniformly selected from 
${l-w+1, \ldots, l} \mod{L}$, then 
($m^{(l_{j})}_{h_{j}},v^{(l_{j})}_{h_{j}}$) is uniformly picked from the 
${\cal P}^{(l_{j})}_{h_{j}}$.

\subsubsection{Variable node operation}
The ($\mlh, \vlh$) $\in {\cal P}^{(l)}_{h}$ for all 
$l \in \{1,\ldots, L\}$ and $h \in \{1,w\}$ is computed by
\begin{align*}
 (\mlh, \vlh) =
 Q \bigl(&(\bar{m}^{(l_{d-1})}_{h_{d-1}},\bar{v}^{(l_{d-1})}_{h_{d-1}}),
 Q \bigl( \cdots, \\
 & Q \bigl( (\bar{m}^{(l_{1})}_{h_{1}},\bar{v}^{(l_{1})}_{h_{1}}),
 (\mu,\sigma^2) \bigr) \cdots \bigr) \bigr) , 
\end{align*}
where $(m,v) = Q\bigl( 
(\bar{m}^{(l_{j})}_{h_{j}}, \bar{v}^{(l_{j})}_{h_{j}}),
(\hat{m}, \hat{v}) \bigr)$ 
is recursively computed as follows.
First $(\hat{m}, \hat{v})$ is initialized by
the channel output $(\mu, \sigma^2)$, then 
\begin{align*}
 m &= \sum_{b\in{\cal B}_{l_{j}}} c'(b) m'(b), \\
 v &= v' + \sum_{b\in{\cal B}_{l_{j}}} c'(b) \bigl(m'(b)\bigr)^2 - \bigl(\sum_{b\in{\cal B}_{l_{j}}} c'(b) m'(b)\bigr)^2,
\end{align*}
where
\begin{align*}
 m'(b) &= v'\bigl( \frac{b}{\bar{v}^{(l_{j})}_{h_{j}} h_j} 
 - \frac{\bar{m}^{(l_{j})}_{h_{j}}}{\bar{v}^{(l_{j})}_{h_{j}}} 
 + \frac{\hat{m}}{\hat{v}} \bigr), \\
 v' &= \frac{\bar{v}^{(l_{j})}_{h_{j}}
 \hat{v}}{\bar{v}^{(l_{j})}_{h_{j}} + \hat{v}}, \\
 c'(b) &= \frac{1}{Z} \exp\bigl( -\frac{(b/h_j-\bar{m}^{(l_{j})}_{h_{j}}
 -\hat{m})^2}{2(\bar{v}^{(l_{j})}_{h_{j}} + \hat{v})}\bigr), \\
 Z &= \sum_{b\in{\cal B}_{l_{j}}} \exp\bigl( -\frac{(b/h_j-\bar{m}^{(l_{j})}_{h_{j}}
 -\hat{m})^2}{2(\bar{v}^{(l_{j})}_{h_{j}} + \hat{v})}\bigr), \\
 {\cal B}_{l_{j}} &= \begin{cases}
    \mathbb{Z} & l_{j} \in \{1,\ldots, L-K+1\}\\
    \{0\} & \text{otherwise}
  \end{cases}.
\end{align*}

Similar to the check node operation, the $d-1$ messages
$(\bar{m}^{(l_{j})}_{h_{j}}, \bar{v}^{(l_{j})}_{h_{j}})$ are chosen as follows:
first the position $l_{j}$ is uniformly selected from 
${l, \ldots, l+w-1} \mod{L}$, then 
$(\bar{m}^{(l_{j})}_{h_{j}}, \bar{v}^{(l_{j})}_{h_{j}})$ is uniformly picked from the 
${\cal \bar{P}}^{(l_{j})}_{h_{j}}$.
In this operation, for each ($\mlh, \vlh$), the $(\mu, \sigma^2)$ is also re-sampled according
to the channel model.

Note that the order of the recursive computation affects the results;
the objective here is to minimize the error due to the single-Gaussian approximation. 
The computations are ordered using the metric
\begin{align*} 
{\cal M}(j) = \frac{(\bar{m}^{(l_{j})}_{h_{j}}+\hat{m})^2}{\bar{v}^{(l_{j})}_{h_{j}} + \hat{v}}.
\end{align*}
We compute $(\bar{m}^{(l_{j})}_{h_{j}}, \bar{v}^{(l_{j})}_{h_{j}})$
beginning with the smallest  ${\cal M}(j)$, 
in order to minimize the single-Gaussian approximation error.

The above procedure repeats until convergence is detected.
The mean of the $v^{(l)}_{w}$ of the message in ${\cal
P}^{l}_{w}$ for all $l \in \{1,\ldots, L\}$ was used to check 
convergence. When the mean (over all $N_{s}$ samples at all positions
$l$) fell below 0.001, 
within $I_{\max}$ iterations, then convergence was declared. 

\section{Experimental Results}
\begin{figure}[t]
  \begin{picture}(200,170)(0,0)
   \put(0,0)
   {
   \put(8,5){\includegraphics[scale=0.32]{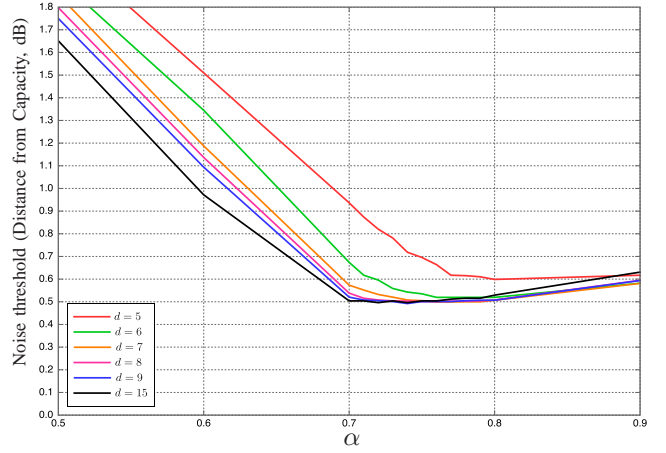}}
   \put(0, 30){\rotatebox{90}{\scriptsize Noise threshold (Distance from
   Capacity, dB)}}
   \put(125, 0){\rotatebox{0}{\normalsize $\alpha$}}
   }
  \end{picture}
   \caption{Noise threshold, measured in distance from capacity, for various
   ($\alpha, d$) LDLC. Average variance of
   messages belonging to nodes labeled $w$ for convergence is $10^{-3}$.
   The number of samples in pool $N_{s}$ is 100000.
   The gap from the capacity is about 0.5 dB.}
   \label{fig:sgde}
 \end{figure}

The noise threshold of conventional ($\alpha, d$) LDLC lattices,
measured in distance from capacity, is shown in Fig.~\ref{fig:sgde}, for $N_{s} = 10^5$ samples. 
The maximum number of iterations is $I_{\max} = 50$.
The noise threshold improves for increasing $\alpha$ and $d$.
In most cases, increasing $\alpha$ above 0.8 had little or no
benefit for improving the threshold. Also, the noise threshold
gradually improves for increasing $d$, however there appears to
be marginal benefit for increasing beyond $d=7$. 
Since the complexity is proportional to $d$,
increasing $d$ beyond this value is not a promising means to improve the threshold.
We observe that the smallest gap from the capacity is about 0.5 dB.

The noise threshold for ($0.8, 7, L, 2$) SC-LDLC lattices proposed in this paper are shown in 
Table~\ref{tab:thVsLwithI}, with $I_{\max}= 5000$ and $20000$, 
for various values of the coupling number $L$. 
We observe that the noise threshold of the (0.8, 7, $L$, 2) SC-LDLC lattice, with
sufficiently large $L$ and $I_{\max}$, is very close to the theoretical limit, within 0.22 dB.
Note that the gap of the capacity is not a precise value, since the capacity
in Eq. \eqref{eqn:cap} assumes lattices defined by full-rank matrices, distinct from 
SC-LDLC lattices. 
The threshold approaches the capacity at small $L$, since the dimension
ratio is small (the capacity in \eqref{eqn:cap} is not valid in this range).
However the capacity loss becomes small, if $L$ becomes large.
For example, the dimension ratio is 0.998 at $L=500$.    
For a practical system which uses shaping, rather than unconstrained power, this is a small penalty.
In addition, the noise threshold appears to converge at $L=15$, 
since the threshold remains 0.22 dB, independently from $L$, for
a sufficiently large number of iterations. 

The noise threshold for various ($\alpha=0.8, d, L, K$) SC-LDLC lattices are shown in 
Table~\ref{tab:incDgrK}, with
$I_{\max}= 5000$. The noise threshold does not improve for increasing
$d$, this is the same observation as in Fig.~\ref{fig:sgde}, 
however slightly improves for increasing $K$ at the same $R_{L}$.
This is because there might be some {\it wiggle} \cite{kudekar2011it}.
There might be a gap from capacity even if the wiggle vanishes with large $K$.

\begin{figure}[htb]
  \begin{center}
   \includegraphics[scale=0.72]{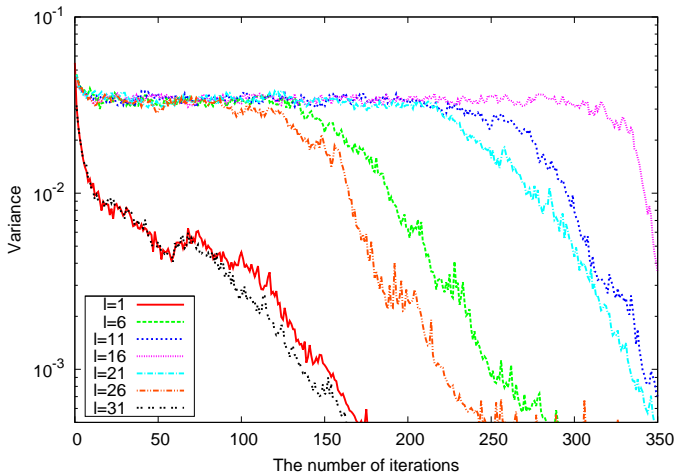}
   \caption{$(\alpha=0.8, d=7, L=31,K=2)$ SC-LDLC, $\sigma^2=0.0548905$: Density evolution
   variance $v^{(l)}_{w}$ at the positions $l=$1, 6, 11, 16, 21, 26, and
   31 at the iterations.}
   \label{fig:mesg}
  \end{center}
 \end{figure}

Fig. \ref{fig:mesg} shows the trajectory of $v^{(l)}_{w}$ at various node
positions as the iterations progress.
As expected, the variances from the variable nodes at the start and end position 
decrease rapidly due to the reliable message from the null check node. 
This phenomenon is the same as the observation in 
spatially-coupled LDPC codes over BEC \cite{kudekar2011it}.

\begin{table}[t]
 \caption{
 Noise threshold (in dB) for various ($\alpha=0.8, d=7, L, K=2$) SC-LDLC lattices
 with $N_{s}=1000$. 
 %It can be observed that SC-LDLC with large $L$ needs large number of iterations. 
 Note that the threshold approaches the capacity at small $L$, since the dimension
 ratio is too small (the capacity in \eqref{eqn:cap} is not valid).
 }
 \label{tab:thVsLwithI}
\begin{center}
  \begin{tabular}{c|ccccccc}
   $L$               & 5    & 8    & 15   & 100  & 300  & 500  \\\hline
   $I_{\max}=5000$   & 0.00 & 0.17 & 0.22 & 0.22 & 0.26 & 0.29 \\
   $I_{\max}=20000$  & 0.00 & 0.17 & 0.22 & 0.22 & 0.22 & 0.22 \\
 \end{tabular}
\end{center}
\end{table}
\begin{table}[t]
 \caption{
 Noise threshold (in dB) for various ($\alpha=0.8, d, L, K$) SC-LDLC with
 $I_{\max}= 5000$ and $N_{s}=1000$. 
 Noise threshold does not improve for increasing $d$, however 
 slightly improves for increasing $K$ at the same $R_{L}$.
 This is because there might be some wiggle \cite{kudekar2011it}.
 }
 \label{tab:incDgrK}
\begin{center}
  \begin{tabular}{c|ccc}
   $d$                      & 7    & 10   & 15   \\\hline
   $L=100, K=2, R_{L}=0.99$ & 0.22 & 0.21 & 0.22 \\
   $L=100, K=5, R_{L}=0.96$ & 0.21 & 0.19 & 0.22 \\
   $L=25,  K=2, R_{L}=0.96$ & 0.22 & 0.21 & 0.21 \\
 \end{tabular}
\end{center}
\end{table}

\section{Conclusion}

This paper has described a new LDLC lattice construction, based upon spatial coupling principles.   Evaluation was performed using Monte Carlo density evolution using a single-Gaussian approximation of the belief-propagation method.   While the conventional lattice construction has a gap of 0.5 dB to capacity, the proposed SC-LDLC construction has a gap of 0.22 dB to capacity, of the unconstrained power channel.

A significant open question remains: how to close up this 0.22 dB gap to capacity?  While spatial coupling improves the noise threshold, the ultimate lattice performance can be no better than the ML performance of the lattice itself.   This opens the opportunity for new code designs, for example, by more closely considering the edge label values, or allowing for non-uniform degree distributions.  On the other hand, we must allow for the possibility that the single-Gaussian approximation introduces errors that limit the threshold under the evaluation method used here.

%\bibliographystyle{IEEEtran}
\bibliographystyle{hieeetr}
\bibliography{gotz-ref}

\begin{thebibliography}{10}

\bibitem{kudekar2011it}
S.~Kudekar, T.~Richardson, and R.~Urbanke, ``Threshold saturation via spatial
  coupling: Why convolutional {LDPC} ensembles perform so well over the
  {BEC},'' {\em IEEE Transactions on Information Theory}, vol.~57,
  pp.~803--834, February 2011.

\bibitem{kudekar2010bms}
S.~Kudekar, C.~M{\'e}asson, T.~J. Richardson, and R.~Urbanke, ``Threshold
  saturation on {BMS} channels via spatial coupling,'' in {\em The 6th
  International Symposium on Turbo Codes and Related Topics}, pp.~319--323,
  September 2010.
\newblock Brest France.

\bibitem{kudekar2011isit-mac}
S.~Kudekar and K.~Kasai, ``Spatially coupled codes over the multiple access
  channel,'' in {\em Proc. 2011 {IEEE} Int. Symp. Inf. Theory (ISIT)}, August
  2011.
\newblock Saint-Petersburg, Russia.

\bibitem{uchi2011isit}
H.~Uchikawa, K.~Kasai, and K.~Sakaniwa, ``Spatially coupled protograph-based
  {LDPC} codes for decode-and-forward in erasure relay channel,'' in {\em Proc.
  2011 {IEEE} Int. Symp. Inf. Theory (ISIT)}, August 2011.
\newblock Saint-Petersburg, Russia.

\bibitem{kudekar2011isit-mem}
S.~Kudekar and K.~Kasai, ``Threshold saturation on channels with memory via
  spatial coupling,'' in {\em Proc. 2011 {IEEE} Int. Symp. Inf. Theory (ISIT)},
  August 2011.
\newblock Saint-Petersburg, Russia.

\bibitem{kudekar2010cs}
S.~Kudekar and H.~D. Pfister, ``The effect of spatial coupling on compressive
  sensing,'' in {\em Proc. 48th Annual Allerton Conf. on Commun., Control and
  Computing}, June 2010.
\newblock Monticello, IL, USA.

\bibitem{takeuchi2011isit}
K.~Takeuchi, T.~Tanaka, and T.~Kawabata, ``Improvement of {BP}-based {CDMA}
  multiuser detection by spatial coupling,'' in {\em Proc. 2011 {IEEE} Int.
  Symp. Inf. Theory (ISIT)}, August 2011.
\newblock Saint-Petersburg, Russia.

\bibitem{hassani2010itw}
S.~H. Hassani, N.~Macris, and R.~Urbanke, ``Coupled graphical models and their
  thresholds,'' in {\em Proc. 2010 {IEEE} Information Theory Workshop (ITW)},
  pp.~1--5, September 2010.

\bibitem{sommer2008it}
N.~Sommer, M.~Feder, and O.~Shalvi, ``Low-density lattice codes,'' {\em IEEE
  Transactions on Information Theory}, vol.~54, pp.~1561--1585, April 2008.

\bibitem{Po1994it}
G.~Poltyrev, ``On coding without restrictions for the {AWGN} channel,'' {\em
  IEEE Transactions on Information Theory}, vol.~40, pp.~409--417, March 1994.

\bibitem{thorpe-JPL2003}
J.~Thorpe, ``{Low} {Density} {Parity} {Check} {(LDPC)} {Codes} {Constructed}
  from {Protographs},'' {\em JPL IPN Progress Report 42-154}, August 2003.

\bibitem{richardson01capacity}
T.~J. Richardson and R.~Urbanke, ``The capacity of low-density parity-check
  codes under message-passing decoding,'' {\em IEEE Trans. on Inform. Theory},
  vol.~47, 2001.

\bibitem{brian09sgde}
B.~M. Kurkoski, K.~Yamaguchi, and K.~Kobayashi, ``Single-gaussian messages and
  noise thresholds for decoding low-density lattice codes,'' in {\em Proc. 2009
  {IEEE} Int. Symp. Inf. Theory (ISIT)}, pp.~734--738, July 2009.

\bibitem{golgo2010mcde}
M.~Gorgoglione, V.~Savin, and D.~Declercq, ``Split-extended {LDPC} codes for
  coded cooperation,'' in {\em Proc. Int. Symp. on Inf. Theory and its
  Applications (ISITA2010)}, pp.~400--405, 2010.

\end{thebibliography}

\end{document}